\documentclass[aps,pra,twocolumn,showpacs,superscriptaddress,letterpaper,longbibliography]{revtex4-1}

\usepackage{graphicx}
\usepackage{mdframed}
\usepackage{framed}
\usepackage{indentfirst}
\usepackage{natbib}
\usepackage{amsmath,amssymb}
\usepackage{subfigure}
\usepackage{enumitem}
\usepackage{epstopdf}
\usepackage{xfrac}
\usepackage{multirow}

\begin{document}

\title{Developing skills vs reinforcing concepts in physics labs: \\Insight from E-CLASS}


\author{Bethany R. Wilcox}
\affiliation{Department of Physics, University of Colorado, 390 UCB, Boulder, CO 80309}

\author{H. J. Lewandowski}
\affiliation{Department of Physics, University of Colorado, 390 UCB, Boulder, CO 80309}
\affiliation{JILA, National Institute of Standards and Technology and University of Colorado, Boulder, CO 80309}

\begin{abstract}
Physics laboratory courses have been generally acknowledged as an important component of the undergraduate curriculum, particularly with respect to developing students' interest in, and understanding of, experimental physics.  There are a number of possible learning goals for these courses including reinforcing physics concepts, developing laboratory skills, and promoting expert-like beliefs about the nature of experimental physics.  However, there is little consensus among instructors and researchers interested in the laboratory learning environment as to relative importance of these various learning goals.  Here, we contribute data to this debate through the analysis of students' responses to the laboratory-focused assessment known as the Colorado Learning Attitudes about Science Survey for Experimental Physics (E-CLASS).  Using a large, national data set of students' responses, we compare students' E-CLASS performance in classes in which the instructor self-reported focusing on developing skills, reinforcing concepts, or both.  As the classification of courses was based on instructor self-report, we also provide additional description of these course with respect to how often students engage in particular activities in the lab.  We find that courses that focus specifically on developing lab skills have more expert-like postinstruction E-CLASS responses than courses that focus either on reinforcing physics concepts or on both goals.  Within first-year courses, this effect is larger for women.  Moreover, these findings hold when controlling for the variance in postinstruction scores that is associated with preinstruction E-CLASS score, student major, and student gender.  
\end{abstract}

\maketitle

\section{\label{sec:intro}Introduction}

Recent years have seen an increase in national calls to study and improve students' undergraduate laboratory course experiences \cite{singer2012dber, olson2012excel}.  In particular, laboratory courses are often called out as a critical component of the undergraduate curriculum with respect to increasing students' understanding of, and interest in, physics.  Despite the general agreement as to the potential value of laboratory courses, there are many possible, and often disparate, learning goals for these courses \cite{singer2006labs, AAPT2015guidelines, trumper2003labs}; too many, in fact, for all to be accomplished in a single course.  Given the variety of potential goals and limited class time, it is nearly always necessary for a laboratory instructors to select a subset of these goals on which to focus.  However, there is little consensus within the literature as to how much emphasis should be placed on each potential goal \cite{singer2006labs}.  

There are at least three general categories of goals that are typically discussed in the literature around laboratory courses: (1) to reinforce the physics concepts taught within the lecture courses \cite{wieman2015labs, singer2006labs, millar2004labs}; (2) to develop students' practical lab skills \cite{AAPT2015guidelines, zwickl2013adlab}; and (3) to foster students' understanding of, and appreciation for, the nature and importance of science generally and physics specifically \cite{trumper2003labs, zwickl2013adlab, millar2004labs, singer2006labs}.  These goals are not, by any means, mutually exclusive; however, the types of activities and assessments that are most often used to support and assess success with respect to these goals can be quite different.  Moreover, these learning goals may not be independent, meaning that a focus on one goal may help or hinder the achievement of one of the other two.  

With respect to the first two goals, the research literature around lab courses typically focuses only on one goal or the other, but not both.  This suggests a potential divide within the community of instructors and researchers working in the laboratory learning environment as to whether lab courses should focus primarily on reinforcing physics concepts or developing lab skills.  In courses where reinforcing concepts is the primary focus, laboratory course success is determined by students scores on test of conceptual mastery of the subject matter (e.g., Refs.\ \cite{wieman2015labs, white1996labs}).  For courses where skills development is the primary focus, students' performance on tests of conceptual mastery become less important than practical assessments of students' ability to, for example, make measurements and collect and interpret accurate data \cite{singer2006labs, bryce1985labAssessment}.  With a few exceptions \cite{day2011cdpa, pillay2008gum}, research-based assessments of students' laboratory skills are virtually non-existent in the literature around laboratory course instruction.  

Fostering more expert-like epistemologies and beliefs around the nature of experimental physics, however, is often an implicit or explicit goal for instructors independent of whether they advocate for a focus on reinforcing concepts or developing lab skills.  Particularly for non-majors or physics students not involved in undergraduate research, lab courses may be their primary opportunity to gain experience with the process of experimental physics and explore its place within physics as a discipline.  This manuscript explores the intersection between the goal of promoting expert-like attitudes and beliefs about experimental physics with instructors' focus on either developing lab skills or reinforcing physics concepts.  To do this, we utilize a large number of students' responses to a laboratory-focused assessment known as the Colorado Learning Attitudes about Science Survey for Experimental physics (E-CLASS) \cite{zwickl2014eclass}.  

The E-CLASS is a 30 item, Likert-style survey that targets students' epistemologies and expectations about the nature and importance of experimental physics, as well as their affect and confidence when doing physics experiments. The E-CLASS presents students with a set of prompts (e.g., ``A common approach for fixing an experiment is to randomly change things until the problem goes away.'') and asks them to agree or disagree with this statement both from their personal perspective when doing experiments in class and that of a hypothetical experimental physicist. The E-CLASS was developed as part of course transformation efforts in the upper-division laboratory courses at the University of Colorado Boulder (CU) \cite{zwickl2013adlab}. The assessment includes items targeting a wide range of learning goals in part because it was intended to be used in both introductory and advanced lab courses. E-CLASS was validated through student interviews and expert review \cite{zwickl2014eclass}, and was tested for statistical validity and reliability using responses from students at multiple institutions and at multiple course levels \cite{wilcox2016eclass}.  

In this paper, we begin by describing the data sources (Sec.\ \ref{sec:data}) and analysis methods (Sec.\ \ref{sec:analysis}) used in this study.  In order to elaborate on the instructor-reported classifications of individual courses, we then present comparisons of these courses with respect to the frequency with which students are asked to complete certain lab activities (Sec.\ \ref{sec:classificationResults}).  Next, we present our findings with respect to whether students in courses focused on developing lab skills versus reinforcing physics concepts gave more expert-like responses to the E-CLASS (Sec.\ \ref{sec:rawResults}).  In addition to looking at trends in the raw postinstruction scores, we also examine whether these trends persist after controlling for the variance associated with other factors such as preinstruction scores, student major, and student gender (Sec.\ \ref{sec:ancovaResults}).  Finally, we end with a discussion of limitations and future work (Sec.\ \ref{sec:discussion}).

\section{\label{sec:methods}Methods}

In this section, we present the data sources, student and institution demographics, and analysis methods used for this study.  

\subsection{\label{sec:data}Data sources}

All data for this study were collected from undergraduate physics lab courses using the E-CLASS centralized administration system \cite{wilcox2016admin} between 01/2015 and 05/2016.  During this period, we collected student responses from 108 distinct courses at 67 institutions.  These institutions included 2-year ($N_{inst}=3$) and 4-year ($N_{inst}=35$) colleges, as well as masters ($N_{inst}=8$) and Ph.D. ($N_{inst}=21$) granting universities.  In a number of the courses, the E-CLASS was used during multiple semesters, and thus the full data set includes student responses from 147 separate instances of the E-CLASS.  These courses included both first-year (FY) introductory courses ($N_{courses}=71$) and beyond-first-year (BFY) courses ($N_{courses}=76$).  Meta-data for each course were collected through an online Course Information Survey (CIS) in which instructors are asked to report information about their course.  In addition to collecting logistical information (e.g., course start and end dates), the CIS also includes questions about course level, instructional strategies, and course structure.  

In the following analysis, we focus in particular on instructors' responses to two questions on the CIS.  In the first question, instructors were asked the following: ``What is the main purpose of the laboratory component of your course?''  
\begin{enumerate}[noitemsep, nolistsep]
\item Reinforce physics concepts
\item Develop lab skills
\item Both
\end{enumerate}

\noindent Based on their responses to this question, each course was tagged as concepts-focused, skills-focused, or both-focused.  In order to better characterize courses in these self-selected categories, we also examine instructors' responses to a second question.  In this second question, instructors were asked to indicate the frequency with which their students engaged in particular lab activities.  The list of activities provided is given in Table \ref{tab:activities}; for each activity, instructors selected one of five frequency options -- never, rarely, sometimes, often, or always.   

\begin{table}
\caption{List of lab activities presented on the CIS.  Instructors were asked to indicate the frequency (never, rarely, sometimes, often, or always) with with their students engaged in these activities.  } \label{tab:activities}
\begin{ruledtabular}
   \begin{tabular}{ p{2.5cm} p{5.8cm} }
      Category & Activity \\
      \hline      
      Type of Investigation & - Verify known physical principles through experimental tests\\
       & - Discover known physical principles through experimentation\\
      & - Explore questions to which the answer is unknown to the student\\
      Student Agency & - Develop their own research questions\\
      & - Design their own procedures\\
      & - Build their own apparatus\\
      & - Choose their own analysis methods\\
      & - Troubleshoot problems with the setup or apparatus\\
      & - Work in groups with other students\\
      Modeling & - Develop \underline{mathematical} models for the \textbf{system} being studied\\
      & - Develop \underline{conceptual} models for the \textbf{system} being studied\\
      & - Develop \underline{mathematical} models for the \textbf{measurement tools} being used\\
      & - Develop \underline{conceptual} models for the \textbf{measurement tools} being used\\
      & - Use mathematical or conceptual models to make predictions\\
      & - Refine system to reduce uncertainty\\
      & - Calibrate measurement tools\\
      Data analysis & - Quantify uncertainty in a measurement\\
      and Visualization & - Calculate uncertainty using error propagation\\
      & - Use computers to aid with data analysis and visualization\\
      & - Use computers to interface with measurement devices\\
      Communication & - Give oral presentations\\
      & - Write lab reports\\
      & - Maintain lab notebooks\\
      & - Read journal articles\\
   \end{tabular}
\end{ruledtabular}
\end{table}

In each course, students completed the E-CLASS online both pre- and postinstruction, typically during the first and last weeks of the lab section.  Only students with matched pre- and postinstruction responses were included in the final data set.  Students were matched based on ID number or first and last name.  Some students' responses were also dropped from the data set based on their answer to a filtering question designed to eliminate students who had not read the item prompts (for more information see Ref. \cite{wilcox2016eclass}).  The final data set included $N=4915$ valid, matched responses.  This represents a matched response rate of roughly 40\% based on estimates of the total enrollment provided by the instructors at the beginning of the course. This response rate is only an approximation of the true response rate as it was based off instructor estimates and enrollment likely fluctuated over the course of the semester.

\begin{table*}
\caption{Demographic breakdown of the full data set, as well as the subset of courses with a focus on developing skills, reinforcing concepts, or both.  Number of courses refers to the number of separate instances of courses, and percentages represent the percentage of students rather than the percentage of courses.  For Major and Gender demographics, the totals may not sum to 100\% as some students did not complete these questions or selected `Other' as their gender.     }\label{tab:dems}
\begin{ruledtabular}
   \begin{tabular}{ l c c c c c c c c }
       & \multicolumn{2}{c}{N} & \multicolumn{2}{c}{Course Level} & \multicolumn{2}{c}{Gender} & \multicolumn{2}{c}{Major} \\
       & Courses & Students & FY & BFY & Women & Men & Physics & Non-physics  \\
     \hline
     All Courses 	& 147 	& 4915 	& 83\% & 17\% & 40\% & 57\% & 21\% & 78\%  \\
     Skills-focused 	& 50 	& 719 		& 44\% & 56\% & 25\% & 73\% & 54\% & 44\%  \\
     Concepts-focused & 19 & 1142 	& 98\% & 2\%   & 32\% & 65\% & 12\% & 86\%  \\
     Both-focused 	& 78 	& 3054 	& 87\% & 13\% & 47\% & 51\% & 16\% & 83\%  \\
   \end{tabular}
\end{ruledtabular}
\end{table*}  

The postinstruction version of the assessment also included a section in which students reported demographic information including their major and gender.  Students were offered 15 distinct major options to chose from; however, we focus on students’ major as the dichotomous distinction between physics and nonphysics majors.  Here, physics includes both engineering physics and physics majors, while nonphysics includes all other majors, including other science majors, nonscience majors, and students who are open option or undeclared.  While we note that students in different nonphysics majors likely had a wide variety of prior and ongoing lab experiences, it is not possible to clearly characterize the nature of these differences given the large number of courses and institutions in the data set.  Given this, and the physics-specific nature of the E-CLASS, we have chosen to focus our analysis of student major specifically on the difference between physics and nonphysics majors.

Table \ref{tab:dems} reports the breakdown of the full data set with respect to course level, student gender, and student major.  Table \ref{tab:dems} also reports the breakdown of these data across courses that were classified as skills-focused, concepts-focused or both-focused.  Note that skills-focused courses were considerably more likely to be BFY than courses focusing on either concepts or both.  This, combined with the typically smaller class sizes associated with BFY courses, also accounts for the smaller-N in the skills-focused group relative to the concepts-focused group despite a greater number of skills-focused courses.  Moreover, only two BFY courses ($N=26$) reported reinforcing physics content as the main purpose of the lab component.  Thus, BFY courses were most often skills- or both-focused, whereas FY courses were most often concepts- or both-focused.

\subsection{\label{sec:analysis}Analysis}

Numerical scores on each of the E-CLASS items were determined based on the established expertlike response for that item \cite{zwickl2014eclass}.  For each item, the responses `(dis)agree' and `strongly (dis)agree' were collapsed into a single `(dis)agree' category, and students' were awarded $+1$ for favorable (i.e., consistent with experts), $+0$ for neutral, and $-1$ for unfavorable (i.e., inconsistent with experts).  A student's overall E-CLASS score was then given by the sum of their scores on each of the 30 items resulting in a possible range of scores of $[-30,30]$ \cite{wilcox2016eclass}.  The distribution of scores on the E-CLASS is typically non-normal, with students concentrated towards positive scores \cite{wilcox2016eclass,wilcox2016gender}.  To account for this, we determine statistical significance based on the non-parametric Mann-Whitney U test \cite{mann1947mwu} unless otherwise stated.  As a measure of effect size and practical significance, we also report Cohen's $d$ \cite{cohen1988d} in cases where differences between distributions were statistically significant \cite{rodriguez2012equity}. 
 
The E-CLASS targets a range of learning goals \cite{zwickl2013adlab, zwickl2014eclass}, some of which may not be relevant to a specific course, thus we have previously cautioned instructors against focusing exclusively on their students' average overall score when interpreting their results \cite{wilcox2016eclass}.  Instead, we suggest that instructors can and should focus also on individual items that they identify as being most relevant to their particular learning goals.  Here, we provide a breakdown of students' scores by item to facilitate that process.  However, the overall score is still useful in that it provides a continuous variable that offers a holistic view of students' performance on the E-CLASS that can be used to quantitatively examine how that performance varies across subpopulations of students.

Previous work with the E-CLASS has demonstrated that students' responses can, and often do, vary significantly across demographic lines (e.g., physics majors vs. nonphysics majors) \cite{wilcox2016gender, wilcox2016eclass}.  Moreover, we have found that demographic differences between subpopulations can confound comparisons of students' E-CLASS scores \cite{wilcox2016gender, wilcox2016pedagogy}.  Thus to account for the demographic differences between skills-, concepts-, and both-focused courses highlighted in Table \ref{tab:dems}, we utilize an analysis of covariance (ANCOVA) \cite{wildt1978ancova} in addition to examining students' raw pre- and postinstruction E-CLASS scores.  ANCOVA is a statistical method for comparing the difference between population means after adjusting them to account for the variance associated with other variables.  As with most statistical tests, valid interpretation of the results of an ANCOVA requires that the data satisfy a number of assumptions \cite{wildt1978ancova, day2016gender}.  Tests of the E-CLASS data indicate that they satisfy these assumptions with one exception.  Preinstruction scores (i.e., the covariate) are not independent of the other variables (e.g., gender or major).  Shared variance between the covariate and independent variables is to be expected in any observational study in which randomized assignment to experimental groups was not done or not possible \cite{miller2001ancova}.  Violation of the assumption of covariate independence implies that our results should be interpreted as a lower bound on the relationship between each independent variable and postinstruction E-CLASS score.

\subsection{\label{sec:classificationResults}Characterizing the courses}

Because the classification of individual courses as being focused on skills, concepts, or both was done based on the instructor's self-report, we cannot unambiguously operationalize what it means for a course to be in each category.  To address this limitation, this section utilizes data from another question on the CIS in order to observationally characterize these different types of courses.   

With respect to how often instructors reported that their students engaged in particular lab activities (Table \ref{tab:activities}), there was significant variation in the responses in courses with different focuses.  For the purposes of statistical comparison, we limit ourselves to comparing only courses identified as being skills-focused or concepts-focused ($N_{courses}=69$) and do not include those courses that were reported as focusing on both concepts and skills ($N_{courses}=78$).  However, we note that for  nearly all activities questions, the average frequency reported by instructors in both-focused courses fell between the averages for the skills- and concepts-focused courses, which is conceptually consistent with our expectations for how these courses might compare.  

Between skills- and concepts-focused courses, there were statistically significant differences (Mann-Whitney U and Holm-Bonferroni corrected $p<0.05$) for one or more items in four of the five categories (see Table \ref{tab:activities}).  With respect to the types of investigations used, instructors in concepts-focused courses reported asking their students to ``verify known physical principles through experimental tests'' more often than instructors in skills-focused courses.  This suggests that skills-focused courses included fewer of the so called ``verification labs.''  In terms of student agency, instructors in skills-focused courses reported asking their students to ``develop their own research questions,'' ``choose their own analysis methods,'' and ``troubleshoot problems with the setup or apparatus'' more often than instructors in concepts-focused courses.  This implies that, overall, skills-focused courses provided more opportunities for students to take agency during lab activities.  In the category of data analysis and visualization, instructors in skills-focused courses reported asking their students to ``quantify uncertainty in a measurement'' more often than those in concepts-focused courses.  There were no statistically significant differences in how often instructors in skills- and concepts-focused courses reported asking their students to engage in particular modeling activities.  

With respect to communication activities, the aggregate data set showed statistically significant differences in the reported frequency for three of the four items -- ``give oral presentations,'' ``maintain lab notebooks,'' and ``read journal articles.''  However, because of the greater representation of BFY courses in the skills-focused group (see Table \ref{tab:dems}), we also looked at comparisons of instructors responses in the FY and BFY courses separately.  The trends were similar for all activity categories except communication.  Separation of the FY and BFY courses showed that BFY instructors in both types of courses were more likely to ask their students to ``give oral presentations'' and ``read journal articles.''  Thus, the apparent differences in instructor responses to these items in skills- and concepts-focused courses were actually artifacts of the differential representation of BFY courses among these two groups.  However, in both FY and BFY courses, skills-focused instructors reported asking their students to ``maintain a lab notebook'' more often than instructors in concepts-focused courses.  

To summarize the trends highlighted in this section, instructors in skills-focused courses used fewer verification labs, provided more opportunities for student agency, and more often asked students to quantify uncertainty in a measurement and maintain a lab notebook.

\section{\label{sec:results}Results}

This section presents findings with respect to whether a focus on skills development or concept reinforcement was accompanied by improvements in students' postinstruction E-CLASS responses using raw scores and an ANCOVA.

\subsection{\label{sec:rawResults}Developing lab skills vs. reinforcing physics content}

\begin{table}[b]
\caption{Overall E-CLASS scores (points) for students in courses focusing on developing skills, reinforcing concepts, or both in the full, aggregate data set ($N=4915$) on both the pre- and post-tests.  ``Sig.'' indicates the statistical significance of the difference between students' scores in courses focusing on skills relative to those focusing on concepts.  \\ * The preinstruction score for both-focused courses was statistically significantly ($p<0.05$) different from the preinstruction scores for either skills-focused or concepts-focused courses both in the FY courses and aggregate data set.   }\label{tab:overall}
\begin{ruledtabular}
   \begin{tabular}{ l l c c c c c}
      Courses & & Skills & Both & Concepts & Sig.  & Effect Size\\
      \hline
      All & N & 719 & 3054 & 1142 & - & - \\
      & Pre & 17.9 & 15.5* & 17.7 &  $p=0.2$ & \\
      & Post & 18.7 & 14.3 & 15.0 & $p\ll0.01$ & $d=0.5$ \\
      FY & N & 316 & 2651 & 1116 & - & - \\
      & Pre & 16.9 & 15.0* & 17.7 &  $p=0.1$ & \\
      & Post & 17.6 & 13.7 & 14.9 & $p\ll0.01$ & $d=0.3$ \\
      BFY & N & 403 & 403 & 26 & - & - \\
      & Pre & 18.7 & 18.2 & 18.5 &  $p=0.9$ & \\
      & Post & 19.6 & 18.2 & 18.2 & $p=0.3$ &  \\
   \end{tabular}
\end{ruledtabular}
\end{table}

\begin{figure}
\includegraphics[width=\linewidth]{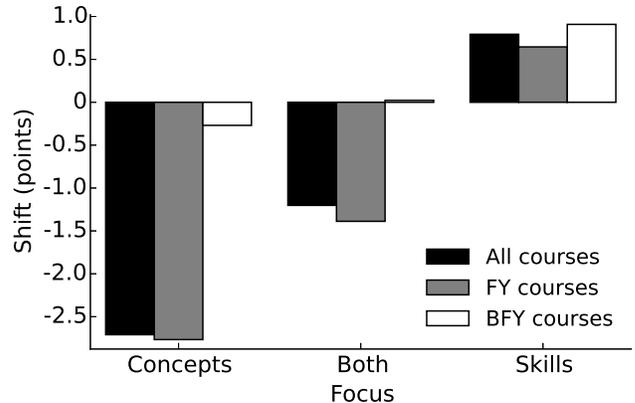}
\caption{Visual representation of pre- to postinstruction shifts in E-CLASS scores from all courses in the data set, as well as for the FY and BFY courses individually.  Differences in the pre- and postinstruction score distributions are statistically significant in all cases except for those of the BFY students in the concepts-focused and both-focused courses. }\label{fig:shifts}
\end{figure}

To identify overall trends in the data, we begin by looking at students' raw overall E-CLASS score both pre- and postinstruction.  Table \ref{tab:overall} reports average scores for all students, and Fig.\ \ref{fig:shifts} offers a visual representation of the shifts in these scores. Because the aggregate trends are dominated by the FY courses, Table \ref{tab:overall} also reports scores for FY and BFY courses separately.  In both the FY and BFY courses, skills-focused courses showed a small ($d=0.1$) but statistically significant ($p\le0.01$) positive shift in overall score.  Concepts-focused courses, on the other hand, showed no shift in BFY courses, and a moderately sized ($d=-0.4$) and statistically significant ($p\ll0.01$) negative shift in FY courses.  FY courses focused on both concepts and skills also showed a statistically significant ($p\ll0.01$) negative shift, but of a somewhat smaller size ($d=-0.2$).  The limited number of BFY, concepts-focused courses ($N=2$) limits our ability to make strong, statistical claims about the comparison between concepts- and skills-focused classes in advanced lab courses.  Because of this, the remainder of our analysis will be restricted to FY courses only ($N=4083$).

\begin{figure*}
\includegraphics[width=\linewidth]{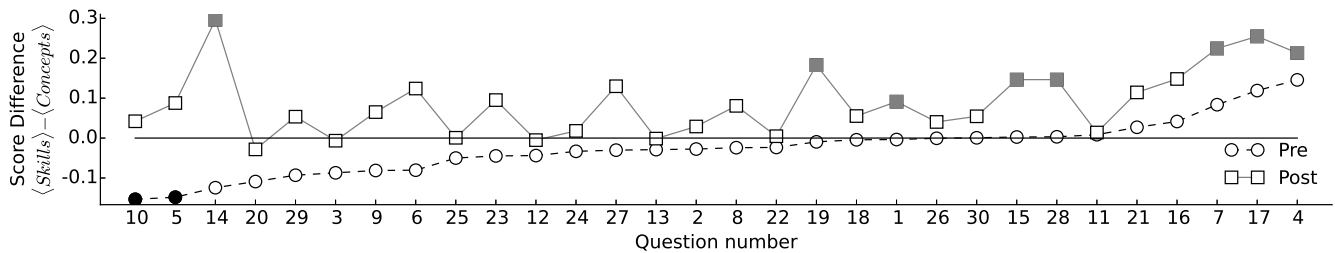}
\caption{Sorted plot of the difference between the average scores (points) of FY students in skills-focused and concepts-focused courses for each item of the E-CLASS. Zero difference is marked by the solid horizontal line.  Filled markers indicate points for which the difference between the distributions was statistically significant (Mann-Whitney U \cite{mann1947mwu} and Holm-Bonferroni \cite{holm1979hb} corrected $p<0.05$).  See Ref.\ \cite{ECLASSwebsite} for full list of item prompts. }\label{fig:byItem}
\end{figure*}

In addition to examining differences in overall score, we also look at students' scores on each of the 30 E-CLASS items.  For the purposes of the by-item comparison, we limit ourselves to comparing only courses identified as being skills-focused or concepts-focused ($N_{courses}=30$) and do not include those courses that were reported as focusing on both concepts and skills ($N_{courses}=41$).  Figure \ref{fig:byItem} shows the difference between the average item scores for FY students in courses with different focuses for both the pre- and postinstruction E-CLASS.  Preinstruction, only two items had statistically significant differences between skills- and concepts-focused courses (Mann-Whitney U \cite{mann1947mwu} and Holm-Bonferroni \cite{holm1979hb} corrected $p<0.05$), and, in both cases, concepts-focused courses scored higher.  Indeed, skills courses had higher scores on only six of the E-CLASS items.  Alternatively, skills-focused courses showed higher scores than concepts-focused courses on all but four items postinstruction.  Differences between postinstruction scores were statistically significant for eight items, with skills-focused courses scoring higher in all cases.  

These findings suggest that students in skills-focused courses tend to have higher postinstruction E-CLASS scores and show more favorable shifts from pre- to postinstruction.  This higher postinstruction performance is driven by small increases in the scores of nearly all items rather than being dominated by large increases in only a few items.  However, Table \ref{tab:dems} also shows significant variation in student demographics between these two types of courses.  Moreover, previous work has shown variations in students' ECLASS responses based on student major and student gender \cite{wilcox2016eclass, wilcox2016gender, wilcox2016structure}.  Previous work has also shown that changes in pedagogy or lab structure can have a differentially positive impact on women \cite{wilcox2016pedagogy}.  The next section addresses these issues using an analysis of covariance.

\subsection{\label{sec:ancovaResults}Analysis of covariance}

The previous section discussed several factors that may be correlated with students' postinstruction E-CLASS responses including course focus (skills vs.\ concepts) and preinstruction score, as well as demographic factors such as student major or gender.  To explore the relationships between these variables simultaneously, we perform an ANCOVA \cite{wildt1978ancova}.  ANCOVA is a statistical method for comparing the difference between population means while adjusting them to account for the variance associated with other variables. In this case, we want to determine if the difference between the E-CLASS scores of students in courses focusing on concepts, skills, or both remains statistically significant after accounting for differences in preinstruction scores, as well as student major and gender.  Only FY students for whom we have matched E-CLASS scores as well as information on both major and gender were included when computing the ANCOVA ($N=3958$).

We performed a 4-way ANCOVA that computed and compared adjusted postinstruction E-CLASS means for courses with different focuses (skills, concepts, or both).  The ANCOVA calculated adjusted means by identifying and accounting for the variance in postinstruction score that was associated with either the covariate (i.e., preinstruction score) or either of the two categorical variables (i.e., student gender or student major).  To account for the possibility that these variables may not be independent of one another, our initial model included all possible interaction terms.  The 4-way ANCOVA revealed a statistically significant interaction between gender and course focus (F-test, $p=0.001$).  The presence of this interaction suggests that the relationship between course focus and postinstruction mean is different for men and women, and implies that the variables gender and course focus must be analyzed independently. A summary of the main findings of the separate ANCOVAs described in the remainder of this section is given in Table \ref{tab:ancova}.

\begin{table}[b]
\caption{Impact of each categorical variable on postinstruction means as adjusted by the 3-way ANCOVAs.  Adjusted means for each variable are calculated controlling for preinstruction score and the other relevant categorical variable, as described in the text.  A difference between group means is indicated only when that difference was statistically significant.  Here, $\langle P\rangle$ is the predicted postinstruction mean for physics students, and similarly for non-physics students $\langle NP\rangle$, men $\langle M\rangle$, women $\langle W \rangle$, skills-focused courses $\langle S\rangle$, concepts-focused courses $\langle C\rangle$, and both-focused courses $\langle B \rangle$.  Statistical significance of the individual comparisons between skills-focused, concepts-focused, and both-focused courses was determined using Tukey's post-hoc test \cite{tukey1949tukey}. } \label{tab:ancova}
\begin{ruledtabular}
   \begin{tabular}{ l c c c}
      & \multicolumn{3}{c}{Catagorical Variable} \\
      Group &  Course Focus & Gender &Major\\
      \hline
      Men      	& $\langle S\rangle>\langle B\rangle>\langle C\rangle$ & - & $\langle P\rangle>\langle NP\rangle$\\
      Women	& $\langle S\rangle>\langle B\rangle>\langle C\rangle$ & - & $\langle P\rangle>\langle NP\rangle$\\
      \hline
      Skills-focused  		& - & $\langle W\rangle>\langle M\rangle$ & $\langle P\rangle>\langle NP\rangle$\\
      Both-focused 		& - & $\langle M\rangle>\langle W\rangle$ & $\langle P\rangle>\langle NP\rangle$ \\
      Concepts-focused 	& - & $\langle W\rangle=\langle M\rangle$ & $\langle P\rangle>\langle NP\rangle$ \\
   \end{tabular}
\end{ruledtabular}
\end{table}

To account for the interaction term, we split the data by gender and ran 3-way ANCOVAs for men and women separately.  Student major was a significant predictor of postinstruction scores for both men and women (F-test, $p<0.01$).  Course focus was also a significant predictor for both men and women (F-test, $p\ll0.01$), with the highest adjusted mean for students in skills-focused courses and the lowest adjusted mean in concepts-focused courses in both cases.  For both men and women, the adjusted means for both-focused courses fell between the means of the skills- and concepts-focused courses, and was statistically significantly different than both (Tukey's post-hoc test \cite{tukey1949tukey}).  However, the increase in postinstruction mean in skills-focused courses was higher for women and the difference was significant (as indicated by the statistical significance of the interaction between gender and course focus).  This finding supports the idea that both men and women scored higher in skills-focused courses, but the difference between skills- and concepts-focused courses was larger for women.  

Similarly, to determine the significance of gender as a predictor of post-instruction E-CLASS scores, we split the data by course focus and ran three separate 3-way ANCOVAs for skills-, concepts-, and both-focused courses separately.  Major was a significant predictor only for students in concepts-focused and both-focused courses (F-test, $p<0.01$), with physics majors scoring higher in both cases.  Gender was a significant predictor of E-CLASS scores only in skills-focused and both-focused courses (F-test, $p<0.01$).  In both-focused courses, men had higher adjusted postinstruction means; however, this trend was reversed for the skills-focused courses, with higher adjusted postinstruction means for women than for men.  This result suggests that the differentially better performance of women in skills-focused courses was enough to eliminate and even reverse the typical gender gap in adjusted postinstruction means.

\section{\label{sec:discussion}Summary and Conclusions}

The success of physics lab courses is often discussed with respect to three potential learning goals: reinforcing physics concepts, developing lab skills, and fostering expert-like beliefs about the nature and importance of experimental physics.  This paper utilized data from a large, national data set of students responses to the E-CLASS to determine the whether the decision to focus primarily on one of the first two learning goals (reinforce concepts or develop skills) was associated with more or less success at achieving the third (fostering expert-like beliefs).   Using instructor self-report, we characterized all courses in our data set as focused on developing skills, reinforcing concepts, or both.  To more clearly characterize differences between the courses in these different categories, we compare them based on how often instructors report asking their students to engage in particular lab activities in their course.  We found that  instructors in skills-focused courses used fewer verification labs, provided more opportunities for student agency, and more often asked students to quantify uncertainty in a measurement and maintain a lab notebook than instructors in concept-focused courses.  

By examining raw E-CLASS scores both overall and by item, we found that students in skills-focused courses showed more expert-like postinstruction responses and more favorable shifts than students in either concepts-focused or both-focused courses.  This result was further supported by an analysis of covariance, which showed that course focus (skills, concepts, or both) was a significant predictor of students postinstruction E-CLASS performance even after adjusting for the variance associated with preinstruction score, student major, and student gender.  Moreover, the ANCOVA demonstrated that the increase in score associated with skills-focused courses was larger for women than for men, and the difference was large enough to eliminate or even reverse the typical gender gap.  Overall, our findings support the conclusion that students in courses that focus primarily on developing lab skills demonstrated greater success with respect to fostering expert-like beliefs about the nature and importance of experimental physics as well as their affect and confidence when doing physics experiments.  

With respect to changes in classroom practice, our finding should be interpreted carefully.  While these results might suggest that improving students' epistemologies, affect, and confidence around experimental physics may be more easily accomplished in courses designed around developing students lab skills, the purely quantitative analysis presented here cannot clearly established a causal connection.  As a possible counterpoint, there is no way to conclusively establish that the higher performance of skills-focused courses relative to concepts-focused courses was not, in part, due to a bias in the instrument.  For example, the types of attitudes and beliefs targeted by the E-CLASS may simply align better with experimental skills learning, while their may be additional items that could be crafted to better align with learning of physics concepts in a lab environment.  However, the E-CLASS was developed based on consensus learning goals developed by practicing experimental physicists \cite{zwickl2014eclass}.  Thus an alternative explanation might be that developing expert-like views about experimental physics requires allowing students to engage authentically with the practice of experimental physics, which in turn requires providing opportunities for them to develop the skills necessary for authentic engagement.  


There are several limitations of this work.  Despite  spanning a large number of institutions, courses, and student populations, our data set is not comprehensive nor randomly selected.  Instructors in our data set generally choose to use E-CLASS without pressure from their department or colleagues, and thus, they represent a self-selected group.  Our data are also skewed towards 4-yr colleges and research universities, with only a few 2-year colleges. We also focused here on a specific subset of potential variables that might confound the comparison of courses with different focuses (i.e., major, gender, and preinstruction scores). We selected these variables based indications from previous work \cite{wilcox2016eclass, wilcox2016gender} that suggested they were important factors in predicting postinstruction E-CLASS scores. Other factors, however, may also be important predictors of students students' E-CLASS responses that may not be independent of an instructors choice of focus (e.g., choice of pedagogy).  Additionally, while we provided some characterization of courses focusing on developing skills or reinforcing concepts, these groups were still based on instructor self-classifications.  We do not have access to information that could provide insight into how a focus on different learning goals actually manifested within the classroom.  Future work could include the inclusion of qualitative data collection through a combination of classroom observations and analysis of classroom artifacts in order to more clearly characterize how instructors target concepts or skills within the lab environment.

\begin{acknowledgments}
This work was funded by the NSF-IUSE Grant No. DUE-1432204 and NSF Grant No. PHY-1125844.  Particular thanks to the members of PER@C for all their help and feedback.  
\end{acknowledgments}

\bibliography{master-refs-ECLASS-10_16}

\begin{thebibliography}{27}%
\makeatletter
\providecommand \@ifxundefined [1]{%
 \@ifx{#1\undefined}
}%
\providecommand \@ifnum [1]{%
 \ifnum #1\expandafter \@firstoftwo
 \else \expandafter \@secondoftwo
 \fi
}%
\providecommand \@ifx [1]{%
 \ifx #1\expandafter \@firstoftwo
 \else \expandafter \@secondoftwo
 \fi
}%
\providecommand \natexlab [1]{#1}%
\providecommand \enquote  [1]{``#1''}%
\providecommand \bibnamefont  [1]{#1}%
\providecommand \bibfnamefont [1]{#1}%
\providecommand \citenamefont [1]{#1}%
\providecommand \href@noop [0]{\@secondoftwo}%
\providecommand \href [0]{\begingroup \@sanitize@url \@href}%
\providecommand \@href[1]{\@@startlink{#1}\@@href}%
\providecommand \@@href[1]{\endgroup#1\@@endlink}%
\providecommand \@sanitize@url [0]{\catcode `\\12\catcode `\$12\catcode
  `\&12\catcode `\#12\catcode `\^12\catcode `\_12\catcode `\%12\relax}%
\providecommand \@@startlink[1]{}%
\providecommand \@@endlink[0]{}%
\providecommand \url  [0]{\begingroup\@sanitize@url \@url }%
\providecommand \@url [1]{\endgroup\@href {#1}{\urlprefix }}%
\providecommand \urlprefix  [0]{URL }%
\providecommand \Eprint [0]{\href }%
\providecommand \doibase [0]{http://dx.doi.org/}%
\providecommand \selectlanguage [0]{\@gobble}%
\providecommand \bibinfo  [0]{\@secondoftwo}%
\providecommand \bibfield  [0]{\@secondoftwo}%
\providecommand \translation [1]{[#1]}%
\providecommand \BibitemOpen [0]{}%
\providecommand \bibitemStop [0]{}%
\providecommand \bibitemNoStop [0]{.\EOS\space}%
\providecommand \EOS [0]{\spacefactor3000\relax}%
\providecommand \BibitemShut  [1]{\csname bibitem#1\endcsname}%
\let\auto@bib@innerbib\@empty
\bibitem [{\citenamefont {Singer}\ \emph {et~al.}(2012)\citenamefont {Singer},
  \citenamefont {Nielsen},\ and\ \citenamefont
  {Schweingruber}}]{singer2012dber}%
  \BibitemOpen
  \bibfield  {author} {\bibinfo {author} {\bibfnamefont {Susan~R}\ \bibnamefont
  {Singer}}, \bibinfo {author} {\bibfnamefont {Natalie~R}\ \bibnamefont
  {Nielsen}}, \ and\ \bibinfo {author} {\bibfnamefont {Heidi~A}\ \bibnamefont
  {Schweingruber}},\ }\href@noop {} {\emph {\bibinfo {title} {Discipline-based
  education research: Understanding and improving learning in undergraduate
  science and engineering}}}\ (\bibinfo  {publisher} {National Academies
  Press},\ \bibinfo {year} {2012})\BibitemShut {NoStop}%
\bibitem [{\citenamefont {Olson}\ and\ \citenamefont
  {Riordan}(2012)}]{olson2012excel}%
  \BibitemOpen
  \bibfield  {author} {\bibinfo {author} {\bibfnamefont {Steve}\ \bibnamefont
  {Olson}}\ and\ \bibinfo {author} {\bibfnamefont {Donna~Gerardi}\ \bibnamefont
  {Riordan}},\ }\bibfield  {title} {\enquote {\bibinfo {title} {Engage to
  excel: Producing one million additional college graduates with degrees in
  science, technology, engineering, and mathematics. report to the
  president.}}\ }\href@noop {} {\bibfield  {journal} {\bibinfo  {journal}
  {Executive Office of the President}\ } (\bibinfo {year} {2012})}\BibitemShut
  {NoStop}%
\bibitem [{\citenamefont {Singer}\ \emph {et~al.}(2006)\citenamefont {Singer},
  \citenamefont {Hilton}, \citenamefont {Schweingruber} \emph
  {et~al.}}]{singer2006labs}%
  \BibitemOpen
  \bibfield  {author} {\bibinfo {author} {\bibfnamefont {Susan~R}\ \bibnamefont
  {Singer}}, \bibinfo {author} {\bibfnamefont {Margaret~L}\ \bibnamefont
  {Hilton}}, \bibinfo {author} {\bibfnamefont {Heidi~A}\ \bibnamefont
  {Schweingruber}},  \emph {et~al.},\ }\href@noop {} {\emph {\bibinfo {title}
  {America's lab report: Investigations in high school science}}}\ (\bibinfo
  {publisher} {National Academies Press},\ \bibinfo {year} {2006})\BibitemShut
  {NoStop}%
\bibitem [{\citenamefont {{AAPT Committee on
  Laboratories}}(2015)}]{AAPT2015guidelines}%
  \BibitemOpen
  \bibfield  {author} {\bibinfo {author} {\bibnamefont {{AAPT Committee on
  Laboratories}}},\ }\href@noop {} {\enquote {\bibinfo {title} {{AAPT
  Recommendations for the Undergraduate Physics Laboratory Curriculum}},}\ }
  (\bibinfo {year} {2015})\BibitemShut {NoStop}%
\bibitem [{\citenamefont {Trumper}(2003)}]{trumper2003labs}%
  \BibitemOpen
  \bibfield  {author} {\bibinfo {author} {\bibfnamefont {Ricardo}\ \bibnamefont
  {Trumper}},\ }\bibfield  {title} {\enquote {\bibinfo {title} {The physics
  laboratory -- a historical overview and future perspectives},}\ }\href
  {\doibase 10.1023/A:1025692409001} {\bibfield  {journal} {\bibinfo  {journal}
  {Science {\&} Education}\ }\textbf {\bibinfo {volume} {12}},\ \bibinfo
  {pages} {645--670} (\bibinfo {year} {2003})}\BibitemShut {NoStop}%
\bibitem [{\citenamefont {Wieman}\ and\ \citenamefont
  {Holmes}(2015)}]{wieman2015labs}%
  \BibitemOpen
  \bibfield  {author} {\bibinfo {author} {\bibfnamefont {Carl}\ \bibnamefont
  {Wieman}}\ and\ \bibinfo {author} {\bibfnamefont {N.~G.}\ \bibnamefont
  {Holmes}},\ }\bibfield  {title} {\enquote {\bibinfo {title} {Measuring the
  impact of an instructional laboratory on the learning of introductory
  physics},}\ }\href@noop {} {\bibfield  {journal} {\bibinfo  {journal}
  {American Journal of Physics}\ }\textbf {\bibinfo {volume} {83}} (\bibinfo
  {year} {2015})}\BibitemShut {NoStop}%
\bibitem [{\citenamefont {Millar}(2004)}]{millar2004labs}%
  \BibitemOpen
  \bibfield  {author} {\bibinfo {author} {\bibfnamefont {Robin}\ \bibnamefont
  {Millar}},\ }\bibfield  {title} {\enquote {\bibinfo {title} {The role of
  practical work in the teaching and learning of science},}\ }\href@noop {}
  {\bibfield  {journal} {\bibinfo  {journal} {High school science laboratories:
  Role and vision}\ } (\bibinfo {year} {2004})}\BibitemShut {NoStop}%
\bibitem [{\citenamefont {Zwickl}\ \emph {et~al.}(2013)\citenamefont {Zwickl},
  \citenamefont {Finkelstein},\ and\ \citenamefont
  {Lewandowski}}]{zwickl2013adlab}%
  \BibitemOpen
  \bibfield  {author} {\bibinfo {author} {\bibfnamefont {Benjamin~M}\
  \bibnamefont {Zwickl}}, \bibinfo {author} {\bibfnamefont {Noah}\ \bibnamefont
  {Finkelstein}}, \ and\ \bibinfo {author} {\bibfnamefont {HJ}~\bibnamefont
  {Lewandowski}},\ }\bibfield  {title} {\enquote {\bibinfo {title} {The process
  of transforming an advanced lab course: Goals, curriculum, and
  assessments},}\ }\href@noop {} {\bibfield  {journal} {\bibinfo  {journal}
  {American Journal of Physics}\ }\textbf {\bibinfo {volume} {81}},\ \bibinfo
  {pages} {63--70} (\bibinfo {year} {2013})}\BibitemShut {NoStop}%
\bibitem [{\citenamefont {White}(1996)}]{white1996labs}%
  \BibitemOpen
  \bibfield  {author} {\bibinfo {author} {\bibfnamefont {Richard~T}\
  \bibnamefont {White}},\ }\bibfield  {title} {\enquote {\bibinfo {title} {The
  link between the laboratory and learning},}\ }\href@noop {} {\bibfield
  {journal} {\bibinfo  {journal} {International Journal of Science Education}\
  }\textbf {\bibinfo {volume} {18}},\ \bibinfo {pages} {761--774} (\bibinfo
  {year} {1996})}\BibitemShut {NoStop}%
\bibitem [{\citenamefont {Bryce}\ and\ \citenamefont
  {Robertson}(1985)}]{bryce1985labAssessment}%
  \BibitemOpen
  \bibfield  {author} {\bibinfo {author} {\bibfnamefont {T.~G.K.}\ \bibnamefont
  {Bryce}}\ and\ \bibinfo {author} {\bibfnamefont {I.~J.}\ \bibnamefont
  {Robertson}},\ }\bibfield  {title} {\enquote {\bibinfo {title} {What can they
  do? a review of practical assessment in science},}\ }\href {\doibase
  10.1080/03057268508559921} {\bibfield  {journal} {\bibinfo  {journal}
  {Studies in Science Education}\ }\textbf {\bibinfo {volume} {12}},\ \bibinfo
  {pages} {1--24} (\bibinfo {year} {1985})}\BibitemShut {NoStop}%
\bibitem [{\citenamefont {Day}\ and\ \citenamefont {Bonn}(2011)}]{day2011cdpa}%
  \BibitemOpen
  \bibfield  {author} {\bibinfo {author} {\bibfnamefont {James}\ \bibnamefont
  {Day}}\ and\ \bibinfo {author} {\bibfnamefont {Doug}\ \bibnamefont {Bonn}},\
  }\bibfield  {title} {\enquote {\bibinfo {title} {Development of the concise
  data processing assessment},}\ }\href {\doibase
  10.1103/PhysRevSTPER.7.010114} {\bibfield  {journal} {\bibinfo  {journal}
  {Phys. Rev. ST Phys. Educ. Res.}\ }\textbf {\bibinfo {volume} {7}},\ \bibinfo
  {pages} {010114} (\bibinfo {year} {2011})}\BibitemShut {NoStop}%
\bibitem [{\citenamefont {Pillay}\ \emph {et~al.}(2008)\citenamefont {Pillay},
  \citenamefont {Buffler}, \citenamefont {Lubben},\ and\ \citenamefont
  {Allie}}]{pillay2008gum}%
  \BibitemOpen
  \bibfield  {author} {\bibinfo {author} {\bibfnamefont {Seshini}\ \bibnamefont
  {Pillay}}, \bibinfo {author} {\bibfnamefont {Andy}\ \bibnamefont {Buffler}},
  \bibinfo {author} {\bibfnamefont {Fred}\ \bibnamefont {Lubben}}, \ and\
  \bibinfo {author} {\bibfnamefont {Saalih}\ \bibnamefont {Allie}},\ }\bibfield
   {title} {\enquote {\bibinfo {title} {Effectiveness of a gum-compliant course
  for teaching measurement in the introductory physics laboratory},}\
  }\href@noop {} {\bibfield  {journal} {\bibinfo  {journal} {European Journal
  of Physics}\ }\textbf {\bibinfo {volume} {29}},\ \bibinfo {pages} {647}
  (\bibinfo {year} {2008})}\BibitemShut {NoStop}%
\bibitem [{\citenamefont {Zwickl}\ \emph {et~al.}(2014)\citenamefont {Zwickl},
  \citenamefont {Hirokawa}, \citenamefont {Finkelstein},\ and\ \citenamefont
  {Lewandowski}}]{zwickl2014eclass}%
  \BibitemOpen
  \bibfield  {author} {\bibinfo {author} {\bibfnamefont {Benjamin~M}\
  \bibnamefont {Zwickl}}, \bibinfo {author} {\bibfnamefont {Takako}\
  \bibnamefont {Hirokawa}}, \bibinfo {author} {\bibfnamefont {Noah}\
  \bibnamefont {Finkelstein}}, \ and\ \bibinfo {author} {\bibfnamefont
  {HJ}~\bibnamefont {Lewandowski}},\ }\bibfield  {title} {\enquote {\bibinfo
  {title} {Epistemology and expectations survey about experimental physics:
  Development and initial results},}\ }\href@noop {} {\bibfield  {journal}
  {\bibinfo  {journal} {Physical Review Special Topics-Physics Education
  Research}\ }\textbf {\bibinfo {volume} {10}},\ \bibinfo {pages} {010120}
  (\bibinfo {year} {2014})}\BibitemShut {NoStop}%
\bibitem [{\citenamefont {Wilcox}\ and\ \citenamefont
  {Lewandowski}(2016{\natexlab{a}})}]{wilcox2016eclass}%
  \BibitemOpen
  \bibfield  {author} {\bibinfo {author} {\bibfnamefont {Bethany~R.}\
  \bibnamefont {Wilcox}}\ and\ \bibinfo {author} {\bibfnamefont {H.~J.}\
  \bibnamefont {Lewandowski}},\ }\bibfield  {title} {\enquote {\bibinfo {title}
  {Students' epistemologies about experimental physics: Validating the colorado
  learning attitudes about science survey for experimental physics},}\ }\href
  {\doibase 10.1103/PhysRevPhysEducRes.12.010123} {\bibfield  {journal}
  {\bibinfo  {journal} {Phys. Rev. Phys. Educ. Res.}\ }\textbf {\bibinfo
  {volume} {12}},\ \bibinfo {pages} {010123} (\bibinfo {year}
  {2016}{\natexlab{a}})}\BibitemShut {NoStop}%
\bibitem [{\citenamefont {Wilcox}\ \emph {et~al.}(2016)\citenamefont {Wilcox},
  \citenamefont {Zwickl}, \citenamefont {Hobbs}, \citenamefont {Aiken},
  \citenamefont {Welch},\ and\ \citenamefont {Lewandowski}}]{wilcox2016admin}%
  \BibitemOpen
  \bibfield  {author} {\bibinfo {author} {\bibfnamefont {Bethany~R.}\
  \bibnamefont {Wilcox}}, \bibinfo {author} {\bibfnamefont {Benjamin~M.}\
  \bibnamefont {Zwickl}}, \bibinfo {author} {\bibfnamefont {Robert~D.}\
  \bibnamefont {Hobbs}}, \bibinfo {author} {\bibfnamefont {John~M.}\
  \bibnamefont {Aiken}}, \bibinfo {author} {\bibfnamefont {Nathan~M.}\
  \bibnamefont {Welch}}, \ and\ \bibinfo {author} {\bibfnamefont {H.~J.}\
  \bibnamefont {Lewandowski}},\ }\bibfield  {title} {\enquote {\bibinfo {title}
  {Alternative model for administration and analysis of research-based
  assessments},}\ }\href {\doibase 10.1103/PhysRevPhysEducRes.12.010139}
  {\bibfield  {journal} {\bibinfo  {journal} {Phys. Rev. Phys. Educ. Res.}\
  }\textbf {\bibinfo {volume} {12}},\ \bibinfo {pages} {010139} (\bibinfo
  {year} {2016})}\BibitemShut {NoStop}%
\bibitem [{\citenamefont {Wilcox}\ and\ \citenamefont
  {Lewandowski}(2016{\natexlab{b}})}]{wilcox2016gender}%
  \BibitemOpen
  \bibfield  {author} {\bibinfo {author} {\bibfnamefont {Bethany~R.}\
  \bibnamefont {Wilcox}}\ and\ \bibinfo {author} {\bibfnamefont {H.~J.}\
  \bibnamefont {Lewandowski}},\ }\bibfield  {title} {\enquote {\bibinfo {title}
  {Research-based assessment of students' beliefs about experimental physics:
  When is gender a factor?}}\ }\href {\doibase
  10.1103/PhysRevPhysEducRes.12.020130} {\bibfield  {journal} {\bibinfo
  {journal} {Phys. Rev. Phys. Educ. Res.}\ }\textbf {\bibinfo {volume} {12}},\
  \bibinfo {pages} {020130} (\bibinfo {year} {2016}{\natexlab{b}})}\BibitemShut
  {NoStop}%
\bibitem [{\citenamefont {Mann}\ and\ \citenamefont
  {Whitney}(1947)}]{mann1947mwu}%
  \BibitemOpen
  \bibfield  {author} {\bibinfo {author} {\bibfnamefont {Henry~B}\ \bibnamefont
  {Mann}}\ and\ \bibinfo {author} {\bibfnamefont {Donald~R}\ \bibnamefont
  {Whitney}},\ }\bibfield  {title} {\enquote {\bibinfo {title} {On a test of
  whether one of two random variables is stochastically larger than the
  other},}\ }\href@noop {} {\bibfield  {journal} {\bibinfo  {journal} {The
  annals of mathematical statistics}\ ,\ \bibinfo {pages} {50--60}} (\bibinfo
  {year} {1947})}\BibitemShut {NoStop}%
\bibitem [{\citenamefont {Cohen}(1988)}]{cohen1988d}%
  \BibitemOpen
  \bibfield  {author} {\bibinfo {author} {\bibfnamefont {J.}~\bibnamefont
  {Cohen}},\ }\href {https://books.google.com/books?id=Tl0N2lRAO9oC} {\emph
  {\bibinfo {title} {Statistical Power Analysis for the Behavioral Sciences}}}\
  (\bibinfo  {publisher} {L. Erlbaum Associates},\ \bibinfo {year}
  {1988})\BibitemShut {NoStop}%
\bibitem [{\citenamefont {Rodriguez}\ \emph {et~al.}(2012)\citenamefont
  {Rodriguez}, \citenamefont {Brewe}, \citenamefont {Sawtelle},\ and\
  \citenamefont {Kramer}}]{rodriguez2012equity}%
  \BibitemOpen
  \bibfield  {author} {\bibinfo {author} {\bibfnamefont {Idaykis}\ \bibnamefont
  {Rodriguez}}, \bibinfo {author} {\bibfnamefont {Eric}\ \bibnamefont {Brewe}},
  \bibinfo {author} {\bibfnamefont {Vashti}\ \bibnamefont {Sawtelle}}, \ and\
  \bibinfo {author} {\bibfnamefont {Laird~H.}\ \bibnamefont {Kramer}},\
  }\bibfield  {title} {\enquote {\bibinfo {title} {Impact of equity models and
  statistical measures on interpretations of educational reform},}\ }\href
  {\doibase 10.1103/PhysRevSTPER.8.020103} {\bibfield  {journal} {\bibinfo
  {journal} {Phys. Rev. ST Phys. Educ. Res.}\ }\textbf {\bibinfo {volume}
  {8}},\ \bibinfo {pages} {020103} (\bibinfo {year} {2012})}\BibitemShut
  {NoStop}%
\bibitem [{\citenamefont {Wilcox}\ and\ \citenamefont
  {Lewandowski}(2016{\natexlab{c}})}]{wilcox2016pedagogy}%
  \BibitemOpen
  \bibfield  {author} {\bibinfo {author} {\bibfnamefont {Bethany}\ \bibnamefont
  {Wilcox}}\ and\ \bibinfo {author} {\bibfnamefont {H.~J.}\ \bibnamefont
  {Lewandowski}},\ }\bibfield  {title} {\enquote {\bibinfo {title} {Impact of
  instructional approach on students' epistemologies about experimental
  physics},}\ }in\ \href@noop {} {\emph {\bibinfo {booktitle} {Physics
  Education Research Conference 2016}}},\ \bibinfo {series} {PER Conference},
  Vol.\ \bibinfo {volume} {Forthcoming}\ (\bibinfo {address} {Sacramento, CA},\
  \bibinfo {year} {2016})\BibitemShut {NoStop}%
\bibitem [{\citenamefont {Wildt}\ and\ \citenamefont
  {Ahtola}(1978)}]{wildt1978ancova}%
  \BibitemOpen
  \bibfield  {author} {\bibinfo {author} {\bibfnamefont {Albert~R}\
  \bibnamefont {Wildt}}\ and\ \bibinfo {author} {\bibfnamefont {Olli}\
  \bibnamefont {Ahtola}},\ }\href@noop {} {\emph {\bibinfo {title} {Analysis of
  covariance}}},\ Vol.~\bibinfo {volume} {12}\ (\bibinfo  {publisher} {Sage},\
  \bibinfo {year} {1978})\BibitemShut {NoStop}%
\bibitem [{\citenamefont {Day}\ \emph {et~al.}(2016)\citenamefont {Day},
  \citenamefont {Stang}, \citenamefont {Holmes}, \citenamefont {Kumar},\ and\
  \citenamefont {Bonn}}]{day2016gender}%
  \BibitemOpen
  \bibfield  {author} {\bibinfo {author} {\bibfnamefont {James}\ \bibnamefont
  {Day}}, \bibinfo {author} {\bibfnamefont {Jared~B.}\ \bibnamefont {Stang}},
  \bibinfo {author} {\bibfnamefont {N.~G.}\ \bibnamefont {Holmes}}, \bibinfo
  {author} {\bibfnamefont {Dhaneesh}\ \bibnamefont {Kumar}}, \ and\ \bibinfo
  {author} {\bibfnamefont {D.~A.}\ \bibnamefont {Bonn}},\ }\bibfield  {title}
  {\enquote {\bibinfo {title} {Gender gaps and gendered action in a first-year
  physics laboratory},}\ }\href {\doibase 10.1103/PhysRevPhysEducRes.12.020104}
  {\bibfield  {journal} {\bibinfo  {journal} {Phys. Rev. Phys. Educ. Res.}\
  }\textbf {\bibinfo {volume} {12}},\ \bibinfo {pages} {020104} (\bibinfo
  {year} {2016})}\BibitemShut {NoStop}%
\bibitem [{\citenamefont {Miller}\ and\ \citenamefont
  {Chapman}(2001)}]{miller2001ancova}%
  \BibitemOpen
  \bibfield  {author} {\bibinfo {author} {\bibfnamefont {Gregory~A}\
  \bibnamefont {Miller}}\ and\ \bibinfo {author} {\bibfnamefont {Jean~P}\
  \bibnamefont {Chapman}},\ }\bibfield  {title} {\enquote {\bibinfo {title}
  {Misunderstanding analysis of covariance.}}\ }\href@noop {} {\bibfield
  {journal} {\bibinfo  {journal} {Journal of abnormal psychology}\ }\textbf
  {\bibinfo {volume} {110}},\ \bibinfo {pages} {40} (\bibinfo {year}
  {2001})}\BibitemShut {NoStop}%
\bibitem [{\citenamefont {Holm}(1979)}]{holm1979hb}%
  \BibitemOpen
  \bibfield  {author} {\bibinfo {author} {\bibfnamefont {Sture}\ \bibnamefont
  {Holm}},\ }\bibfield  {title} {\enquote {\bibinfo {title} {A simple
  sequentially rejective multiple test procedure},}\ }\href@noop {} {\bibfield
  {journal} {\bibinfo  {journal} {Scandinavian journal of statistics}\ ,\
  \bibinfo {pages} {65--70}} (\bibinfo {year} {1979})}\BibitemShut {NoStop}%
\bibitem [{\citenamefont {{tinyurl.com/ECLASS-physics}}(2015)}]{ECLASSwebsite}%
  \BibitemOpen
  \bibfield  {author} {\bibinfo {author} {\bibnamefont
  {{tinyurl.com/ECLASS-physics}}},\ }\href@noop {} {} (\bibinfo {year}
  {2015})\BibitemShut {NoStop}%
\bibitem [{\citenamefont {Wilcox}\ and\ \citenamefont
  {Lewandowski}(2016{\natexlab{d}})}]{wilcox2016structure}%
  \BibitemOpen
  \bibfield  {author} {\bibinfo {author} {\bibfnamefont {Bethany~R.}\
  \bibnamefont {Wilcox}}\ and\ \bibinfo {author} {\bibfnamefont {H.~J.}\
  \bibnamefont {Lewandowski}},\ }\bibfield  {title} {\enquote {\bibinfo {title}
  {Open-ended versus guided laboratory activities:impact on students' beliefs
  about experimental physics},}\ }\href {\doibase
  10.1103/PhysRevPhysEducRes.12.020132} {\bibfield  {journal} {\bibinfo
  {journal} {Phys. Rev. Phys. Educ. Res.}\ }\textbf {\bibinfo {volume} {12}},\
  \bibinfo {pages} {020132} (\bibinfo {year} {2016}{\natexlab{d}})}\BibitemShut
  {NoStop}%
\bibitem [{\citenamefont {Tukey}(1949)}]{tukey1949tukey}%
  \BibitemOpen
  \bibfield  {author} {\bibinfo {author} {\bibfnamefont {John~W.}\ \bibnamefont
  {Tukey}},\ }\bibfield  {title} {\enquote {\bibinfo {title} {Comparing
  individual means in the analysis of variance},}\ }\href
  {http://www.jstor.org/stable/3001913} {\bibfield  {journal} {\bibinfo
  {journal} {Biometrics}\ }\textbf {\bibinfo {volume} {5}},\ \bibinfo {pages}
  {99--114} (\bibinfo {year} {1949})}\BibitemShut {NoStop}%
\end{thebibliography}%

\end{document}